\documentclass[prl,12pt]{revtex4}
\usepackage{amssymb}
\usepackage{amsmath}
\usepackage{graphicx}
\usepackage{color}
\usepackage{tabularx}
\DeclareGraphicsExtensions{.jpg,.jpeg,.pdf,.png,.mps,.eps,.ps}

\newcommand{\beq}{\begin{equation}}
\newcommand{\eeq}{\end{equation}}
\newcommand{\bea}{\begin{eqnarray}}
\newcommand{\ena}{\end{eqnarray}}
\newcommand{\etal}{{\it et al.}}
\newcommand{\ie}{{\it i.e.}}
\newcommand{\eg}{{\it e.g.}}
\newcommand{\etc}{{\it etc.}}
\newcommand{\lsim}{\mathrel{\mathop{\kern 0pt \rlap{\raise.2ex\hbox{$<$}}}\lower.9ex\hbox{\kern-.190em $\sim$}}}
\newcommand{\gsim}{\mathrel{\mathop{\kern 0pt \rlap{\raise.2ex\hbox{$>$}}}\lower.9ex\hbox{\kern-.190em $\sim$}}}

\newcommand{\physics}[1]{{\tt physics/#1}}

\newcommand{\prep}[3]{Phys.\ Rep.\ {\bf #1}, #3 (#2)}
\newcommand{\plb}[3]{Phys.\ Lett.\ B\ {\bf #1}, #3 (#2)}

\newcommand{\pr}[3]{Phys.\ Rev.\ {\bf #1}, #3 (#2)}
\renewcommand{\prl}[3]{Phys.\ Rev.\ Lett. {\bf #1}, #3 (#2)}
\renewcommand{\prd}[3]{Phys.\ Rev.\ D\ {\bf #1}, #3 (#2)}

\renewcommand{\rmp}[3]{Rev.\ Mod.\ Phys.\ {\bf #1}, #3 (#2)}

\newcommand{\href}[2]{#1}

\definecolor{cyan}{cmyk}{1.,0.,0.,0.5}
\definecolor{magenta}{cmyk}{0.,1.,0.,0.5}
\definecolor{verdatre}{cmyk}{0.5,0.,0.5,0.5}
\definecolor{yellow}{cmyk}{0.,0.,0.2,0.0}
\definecolor{rouge}{cmyk}{0.,0.4,0.6,0.0}
\definecolor{orange}{cmyk}{0.,0.5,0.5,0.}
\definecolor{violet}{rgb}{0.5,0.,0.5}

\begin{document}

\noindent
\title{Essence of Special Relativity, Reduced Dirac Equation and Antigravity}
 \vskip 1.cm
\author{Guang-jiong Ni $^{\rm a,b}$}
\email{\ pdx01018@pdx.edu}
\affiliation{$^{\rm a}$ Department of Physics, Portland State University, Portland, OR97207, U. S. A.\\
$^{\rm b}$ Department of Physics, Fudan University, Shanghai, 200433, China}

\author{Suqing Chen $^{\rm b}$}
\email{\ suqing\_chen@yahoo.com}
\affiliation{$^{\rm b}$ Department of Physics, Fudan University, Shanghai, 200433, China}

\author{Senyue Lou $^{\rm c,d}$}
\email{\ sylou@sjtu.edu.cn}
\affiliation{$^{\rm c}$ Department of Physics, Shanghai Jiao Tong University, Shanghai, 200030, China\\
$^{\rm d}$ Department of Physics, Ningbo University, Ningbo 315211, China}

\author{Jianjun Xu $^{\rm b}$}
\email{\ xujj@fudan.edu.cn}
\affiliation{$^{\rm b}$ Department of Physics, Fudan University, Shanghai, 200433, China}

\vskip 0.5cm
\date{\today}

\vskip 0.5cm
\begin{abstract}
The essence of special relativity is hiding in the equal existence of particle and antiparticle, which can be expressed
by two discrete symmetries within one inertial frame --- the invariance under the (newly defined)
space-time inversion (${\bf x}\to -{\bf x},t\to -t$), or equivalently, the invariance under a mass inversion
($m\to -m$). The problems discussed are: the evolution of the $CPT$ invariance into a basic postulate, an unique
solution to the original puzzle in Einstein-Podolsky-Rosen paradox, the reduced Dirac equation for hydrogenlike
atoms, and the negative mass paradox leading to the prediction of antigravity between matter and antimatter.
\end{abstract}

\maketitle

\vskip 1cm
\section{Introduction}\label{sec:introduction}

\vskip 0.1cm

It's time to rethink the theory of special relativity ($SR$) established by Einstein in 1905. As is well-known, he
put $SR$ on two basic postulates:\\
$A$. Principle of constancy of the light speed;\\
$B$. Principle of relativity.

Both $A$ and $B$ are "relativistic principles". Why we need both of them? Some authors have been devoting to reduce these
two principles into one, trying to ignore $A$. They might think as follows: Once we already have Maxwell's theory of classical electrodynamics ($CED$) in one inertial frame $S$, the principle B would lead to the conclusion that the $CED$ must hold in another inertial frame $S'$, \ie, principle $A$ must be valid too.

All attempts mentioned above were doomed to failure because they overlooked one clue point: In order to transfer from
one frame $S$ to another $S'$, we need a transformation bringing the coordinates from ${\bf x},t$ to ${\bf x}',t'$.
And in this Lorentz transformation, an universal constant $c$ must be fixed in advance, otherwise the transformation would be meaningless.

However, the above argument, so-called a "logic cycle", does hint that the $SR$ based on $A$ and $B$ is by no means
a final story. If we insist on discussing physics in one frame, there should be essentially one "relativistic principle"
only. Like everything else, the essence of $SR$ can only be revealed gradually after 1905. But where is the breakthrough point?

\section{The $CPT$ Invariance Turning into a Basic Postulate}\label{sec:invariance}

\vskip 0.1cm

There are three discrete transformations in quantum mechanics ($QM$)\cite{1}, quantum field theory ($QFT$) and particle physics\cite{2}:\\
(a) Space-inversion ($P$):

The sign change of space coordinates (${\bf x}\to -{\bf x}$) in the wavefuction ($WF$) of $QM$ may lead to two eigenstates:
\begin{equation}\label{}
\psi_{\pm}({\bf x},t)\to \psi_{\pm}(-{\bf x},t)=\pm\psi_{\pm}({\bf x},t)
\end{equation}
with eigenvalues $1$ or $-1$ being the even or odd parity.

(b) Time reversal ($T$):

The so-called $T$ transformation is actually not a "time reversal" but a "reversal of motion"\cite{3,4}, which implies an
equivalence relation between two $WFs$:
\begin{equation}\label{}
\psi({\bf x},t)\sim \psi^*({\bf x},-t)
\end{equation}
where the equivalence notation $\sim$ possibly means some matrices in front of $WF$ being ignored.

(c) Charge conjugation transformation($C$):

The $C$ transformation brings a particle (with charge $q$) into its antiparticle (with charge $-q$) and implies
a complex conjugation on the $WF$:
\begin{equation}\label{psic}
\psi({\bf x},t)\to \psi_c({\bf x},t)\sim \psi^*({\bf x},t)
\end{equation}
Note that the $WF$ $\psi_c$ implies a negative-energy particle. To explain it being an antiparticle, one has to resort
to so-called "hole theory" for electron --- the vacuum is fully filled with infinite negative-energy electrons and a
"hole" created in the "sea" would correspond to a positron\cite{1,5}. But how could the "hole theory" be applied to
the boson particle? No one knows. (See the discussion between S. Weinberg and Dirac\cite{6}).

(d) $CPT$ combined transformation

If taking the product of $C$, $P$ and $T$ transformations together, the complex conjugation contained in the $C$ and $T$
will cancel each other, yielding\cite{2,5}
\begin{equation}\label{cpt}
\psi({\bf x},t)\to CPT\psi({\bf x},t)=\psi_{CPT}({\bf x},t)\sim \psi(-{\bf x},-t)
\end{equation}
On the right-hand-side (RHS), the $WF$ should be understood as to describe an antiparticle. But it differs from the original
$WF$ only in the sign change of ${\bf x}$ and $t$. What does it mean?

The historical discovery of parity violation in 1956-1957\cite{7,8} reveals that both $P$ and $C$ symmetries are violated
in weak interactions. Since 1964, it is found that $CP$ symmetry is also violated whereas the $CPT$ invariance remains valid
\cite{2}, which in turn implies the violation of $T$ reversal symmetry, as summarized in the Review of particle Physics\cite{9}.

Therefore, the relation between a particle $|a\rangle$ and its antiparticle $|\bar{a}\rangle$ is not $|\bar{a}\rangle=C|a\rangle$ but (as defined by Lee and Wu\cite{10})
\begin{equation}\label{anti}
|\bar{a}\rangle=CPT|a\rangle
\end{equation}
which means exactly the Eq.(\ref{cpt}). For example, for an electron in free motion, its $WF$ reads
\begin{equation}\label{elec}
\langle{\bf x},t|e^-,{\bf p},E\rangle=\psi_{e^-}({\bf x},t)\sim \exp\left[\frac{i}{\hbar}({\bf p}\cdot{\bf x}-Et)\right]
\end{equation}
while the $WF$ for a positron is given by Eq.(\ref{cpt}) or Eq.(\ref{anti}) as
\begin{equation}\label{posi}
\langle{\bf x},t|e^+,{\bf p},E\rangle=\psi_{e^+}({\bf x},t)\sim \exp\left[-\frac{i}{\hbar}({\bf p}\cdot{\bf x}-Et)\right]
\end{equation}
Note that the momentum ${\bf p}$ and energy $E\,(>0)$ are the same in Eqs.(\ref{elec}) and (\ref{posi}) (see Eq.(16.51)
in \cite{2}).

The above relation should be viewed as a new symmetry: The (newly defined) space-time inversion (${\bf x}\to -{\bf x},t\to -t$) is equivalent to particle-antiparticle transformation. The transformation of a particle to its antiparticle (denoted by
$\cal C$) is not something which can be defined independently but a direct consequence of the (newly defined) space-time
inversion $\cal PT$ (${\bf x}\to -{\bf x},t\to -t$)\cite{11,12,4}:
\begin{equation}\label{ptc}
{\cal PT}={\cal C}
\end{equation}
Note that there is an important difference between a "theorem" and a "law". Various quantities contained in a theorem
must be defined clearly and unambiguously in advance before the theorem can be proved. On the other hand, a law can often
(not always) accommodate a definition of a physical quantity which can only be defined unambiguously after the law is verified by experiments. Two examples are:

The definition of inertial mass $m$ is contained in the Newton's dynamical law:
\begin{equation}\label{newt}
{\bf F}=m{\bf a}
\end{equation}
The definition of electric (magnetic) field strength $\bf E\,(B)$ is contained in the Lorentz-force formula:
\begin{equation}\label{loren}
{\bf F}=m\dfrac{d{\bf v}}{dt}=q({\bf E}+\frac{1}{c}{\bf v}\times{\bf B})
\end{equation}
Hence, we see from Eqs.(\ref{elec}) and (\ref{posi}) that the familiar operator relations in $QM$:
\begin{equation}\label{pe}
\hat{\bf p}=-i\hbar\nabla,\quad \hat{E}=i\hbar\frac{\partial}{\partial t}
\end{equation}
are only valid for particle, they must be supplemented by
\begin{equation}\label{pec}
\hat{\bf p}_c=i\hbar\nabla,\quad \hat{E}_c=-i\hbar\frac{\partial}{\partial t}
\end{equation}
for antiparticle, in conformity with the basic postulate, Eq.(\ref{ptc}).

From the beginning, we have been believing that Eq.(\ref{ptc}) or Eq.(\ref{pe}) versus Eq.(\ref{pec}) is the essence of $SR$. We derived $SR$ from $QM$, based on this symmetry\cite{12}. Among various arguments for this claim (\cite{13}, see \cite{4} for detail), we will discuss the $EPR$ paradox in the next section before turning to another new arguments in our recent studies.

\section{The Original Puzzle in Einstein-Podolsky-Rosen Paradox}
\label{sec:paradox}

\vskip 0.1cm

The famous paper titled "Can quantum mechanical description of physical reality be considered complete?" by
Einstein, Podolsky and Rosen ($EPR$, \cite{14}) is not easy to read. Quite naturally, beginning from Bohm \cite{15}
and Bell \cite{16}, physicists have been turning their attention to the entanglement phenomena of photons and electrons with spin. To our knowledge, H. Guan (1935-2007) first clearly pointed out that \cite{17} the original puzzle in $EPR$'s paper is involving spinless particles and what being overlooked is as follows.

Consider two particles in one dimensional space with positions $x_i,\,(i=1,2)$ and momentum operators $\hat{p}_i=-i\hbar\dfrac{\partial}{\partial x_i}$. Then the commutation relation
\begin{equation}\label{xps}
[x_1-x_2,\hat{p}_1+\hat{p}_2]=0
\end{equation}
implies that there may be a state with two commutative  (compatible) observables:
\begin{equation}\label{xps1}
p_1+p_2=0,\quad (p_2=-p_1)\quad\text{and}\quad x_1-x_2=D
\end{equation}
How can such a quantum state be realized?

Guan's observation led to discussions in Refs.\cite{18}, \cite{19} and \cite{4}, where another commutation relations like
\begin{eqnarray}
  {[x_1+x_2,\hat{p}_1-\hat{p}_2]} &=& 0 \label{xps2}\\
  {[t_1-t_2,\hat{E}_1+\hat{E}_2]} &=& 0 \label{xps3}\\
  {[t_1+t_2,\hat{E}_1-\hat{E}_2]} &=& 0\label{xps4}
\end{eqnarray}
($\hat{E}_i=i\hbar\frac{\partial}{\partial t_i}$) are considered in connection with a wonderful experiment (in 1998) on
an entangled state of $K^0-\bar{K}^0$ system \cite{20}. Now let us discuss it further.

As in \cite{20}, we focus on back-to-back events. However, the evolution of wavefunctions ($WFs$) will be considered in three inertial frames: The center-of-mass system $S$ is at rest with its origin $t=0$ located at detector's center.
The space-time coordinates in Eqs.(\ref{xps})-(\ref{xps4}) refer to particles moving to right ($x_1>0$) and left ($x_2<0$) respectively.
Then we take an inertial system $S'$ with its origin located at particle $1$ (\ie, $x'_1=0$). $S'$ is moving in a uniform
velocity $v$ with respect to $S$. (For Kaon's momentum of $800 MeV/c, \beta=\frac{v}{c}=0.849$). Another $S''$ system is
chosen with its origin located at particle $2$ ($x''_2=0$). $S''$ is moving in a velocity $-v$ with respect to $S$. Thus we
have Lorentz transformations among their space-time coordinates as:
\begin{equation}\label{lorentz}
\left\{
  \begin{array}{ll}
    x'=\dfrac{x-vt}{\sqrt{1-\beta^2}}, & \\
    t'=\dfrac{t-vx/c^2}{\sqrt{1-\beta^2}}, &
  \end{array}
\right.\qquad
\left\{
  \begin{array}{ll}
    x''=\dfrac{x+vt}{\sqrt{1-\beta^2}}, & \\
    t''=\dfrac{t+vx/c^2}{\sqrt{1-\beta^2}}, &
  \end{array}\right.
\end{equation}
Here $t'_1$ and $t''_2$ correspond to the proper times $t_a$ and $t_b$ in \cite{20} respectively. The common time origin
$t=t'=t''=0$ is adopted.

Surprisingly, we see that Eqs.(\ref{xps}) and (\ref{xps1}) are just realized by $K^0K^0$ events with $p_2=-p_1$ and $x_1-x_2=D$
being the distance between two particles when they are detected. Note that Eq.(\ref{xps4}) is also realized in this case
with $E_1=E_2$ and $t_1+t_2=D/v$.

More interestingly, Eqs.(\ref{xps2}) and (\ref{xps3}) can be realized exactly by $K^0\bar{K}^0$ events. For example, if particle
$1$ is $K^0$, then particle $2$ must be a $\bar{K}^0$ with ${\hat p}_2=-{\hat p}_2^c$ and ${\hat E}_2=-{\hat E}_2^c$ (see
Eq.(\ref{pec})). So $p_1=-p_2^c>0, E_1=E_2^c>0$ and $x_1+x_2=v(t_1-t_2)=\Delta l$ implies the flight-path difference measured in the laboratory \cite{20}.

A remarkable merit of experiment in \cite{20} lies in the fact that its data cover $t_1\neq t_2$ cases and so go
beyond the EPR-type correlation ($t_1=t_2$). However, the concept of "simultaneity" in time is relative and frame dependent as can be seen from Eq.(\ref{lorentz}):
\begin{eqnarray}
  t'_2-t'_1 &=& \frac{1}{\sqrt{1-\beta^2}}[(t_2-t_1)+\beta^2(t_1+t_2)]>0, \quad (t_2>t_1) \label{19}\\
  t''_1-t''_2 &=& \frac{1}{\sqrt{1-\beta^2}}[(t_1-t_2)+\beta^2(t_1+t_2)]>0, \quad (t_1>t_2)\label{20}
\end{eqnarray}
Here $t'_1$ or $t''_2$ is the proper time of particle first observed in the $K^0-\bar{K}^0$
system. We see that the "causality" is preserved because even at the EPR limit ($t_1=t_2$), Eqs.(\ref{19}) and (\ref{20}) remain positive.

Once a particle (say $2$) is first detected, a destruction process on the coherence of entangled state is triggered. This process will be accomplished right at the detection of second particle (say $1$). In order to better understand why the coherence can be maintained within this interval, let us stay at $S''$ system and compare two velocities. The particle $1$ has a velocity being
\begin{equation}\label{velo1}
v''_1=\frac{x''_1}{t''_1}=\frac{2v}{1+v^2/c^2}
\end{equation}
as expected. On the other hand, the correlation between particles
$1$ and $2$ has been established since $t''=0$ until $t''_2$, during which a
"decoherence signal" from particle $2$ is triggered and it reaches
particle $1$ at time $t''_1$. The signal's propagation velocity must be no less than
\begin{equation}\label{}
w''=\frac{x''_1-x''_2}{t''_1-t''_2}=\frac{2v}{(1+\beta^2)-(1-\beta^2)t_2/t_1}\,\xrightarrow[t_1\to t_2]\,\frac{c^2}{v}>c
\end{equation}
which is superluminal! However, we learn from $RQM$ that the wave's phase velocity $u_p=\frac{\omega}{k}=\frac{E}{p}$,
($E^2=p^2c^2+m^2c^4$) is different from its group
velocity  $u_g$, \ie, the particle's velocity $v$ ($v=u_g=\frac{d\omega}{dk}=\frac{pc^2}{E}$) with their relation being:
\begin{equation}\label{velo}
u_pu_g=c^2,\quad u_p=c^2/v
\end{equation}
so $u''_p =c^2/v''|_{v''\to 0}\to \infty$ as measured from particle $2$. The inequality
$v''_1<w'' <u''_p$ ensures the quantum correlation surviving throughout
the time interval $t''_2< t'' <t''_1$.

The phase velocity is by no means an observable speed of energy
transfer, but it does keep the wave coherence globally. By contrast,
the destruction of coherence is triggered and accomplished by
detectors locally. In the wavefunction ($WF$) of $K-K$ system, there
are both $K^0$ and $\bar{K}^0$ (with their space-time coordinates) in the wave
propagating to the right or left side. Actually, there is neither $K^0$
nor $\bar{K}^0$ particle in the wave but a wave with its interference until
particles $1$ and $2$ are detected eventually with only one particle (
either $K^0$ or $\bar{K}^0$ ) at each side \cite{20}. In our understanding \cite{4}, the
invisible $WF$ is the amplitude of a "fictitious measurement" to show
the relevant "potential possibility", which turns into the "real
probability" at a concrete measurement.

Hence $EPR$ were right: For better understanding the $QM$, one needs to
study the two-particle entangled state, a nonlocal coherent state
evolved over long distance not only in space but also in time. As
shown by experiment \cite{20} and Eqs.(\ref{xps})-(\ref{xps4}), being a quantum system with less
uncertainty, it is easier to be observed. This is because anything
could be and should be recognized only in relationships as emphasized
by $SR$.

$EPR$ were quite right: $QM$ cannot be considered complete in
describing the physical reality unless (a). we take the antiparticle with
relevant operator relations, Eq.(\ref{pec}), into account: (b). the
relativistic relation between phase velocity and group velocity, Eq.
(\ref{velo}), is also taken into account. \footnotemark[1]
\footnotetext[1]{If constrained by nonrelativistic relation $E= p^2/(2m)$ which would
lead to $u_p=(1/2)u_g= (1/2) v$, we even cannot understand the forming
process toward the Bose-Einstein-condensation where any two particles
obey Eqs.(\ref{xps})-(\ref{xps4}).}

\section{Invariance Under Mass Inversion and Dirac Equation}
\label{sec:massinversion}

\vskip 0.1cm

There is another symmetry equivalent to the invariance of space-time inversion, Eq.(\ref{ptc}), showing the equal
existence of particle and antiparticle.

As noticed in \cite{4}, the Lorentz force law, Eq.(\ref{loren}) for an electron should be transformed into that for
a positron by an inversion of $m\to -m$ as a substitution of changing $q=-e\,(e>0)$ into $q=e$ in classical physics.
This is because the particle-antiparticle transformation has already been replaced by Eq.(\ref{anti}) or Eq.(\ref{ptc}).
Moreover, we can see from Eqs.(\ref{elec}) and (\ref{posi}) that the space-time inversion (${\bf x}\to -{\bf x},t\to -t$) is equivalent to changing the sign of the $\bf p$ and $E$, \ie, changing $m\to -m$.

However, it was not until 2003 that the importance of mass inversion became clearer as discussed in \cite{21} (see also
Appendix 9C in the 2nd Edition of \cite{4}). Let us look at the Dirac equation for a free electron
\begin{equation}\label{dirac}
i\hbar\dfrac{\partial}{\partial t}\psi=H\psi=(-i\hbar c{\boldsymbol\alpha}\cdot\nabla +\beta mc^2)\psi
\end{equation}
with ${\boldsymbol\alpha}$ and $\beta$ being $4\times4$ matrices, the $WF$ $\psi$ is a four-component spinor:
\begin{equation}\label{2spin}
\psi=\begin{pmatrix}\phi \\ \chi\end{pmatrix}
\end{equation}
Usually, the two-component spinors $\phi$ and $\chi$ are called "positive" and "negative" energy components. In our point of
view, they are the hiding "particle" and "antiparticle" fields in a particle (electron) (\cite{4}, see below). Substitution of Eq.(\ref{2spin}) into Eq.(\ref{dirac}) leads to
\begin{equation}\label{dirac2}
\left\{\begin{array}{l}
i\hbar\dfrac{\partial}{\partial t}\phi=-i\hbar c{\boldsymbol\sigma}\cdot\nabla\chi+mc^2\phi, \\
i\hbar\dfrac{\partial}{\partial t}\chi=-i\hbar c{\boldsymbol\sigma}\cdot\nabla\phi-mc^2\chi
      \end{array}\right.
\end{equation}
(${\boldsymbol\sigma}$ are Pauli matrices). Eq.(\ref{dirac2}) is invariant under the space-time inversion (${\bf x}\to -{\bf x},t\to -t$) with
\begin{equation}\label{inver}
\phi(-{\bf x},-t)\to\chi({\bf x},t),\quad \chi(-{\bf x},-t)\to\phi({\bf x},t)
\end{equation}
Alternatively, it also remains invariant under a mass inversion as (see also \cite{19})
\begin{equation}\label{}
m\to -m,\quad \phi({\bf x},t)\to\chi({\bf x},t),\quad \chi({\bf x},t)\to\phi({\bf x},t)
\end{equation}
Note that the transformation $m\to -m$ by no means implies antiparticle having "negative mass". Both particle and antiparticle have positive mass as shown by using Eqs.(\ref{pe}) and (\ref{pec}) respectively. This will be clearer later.

\section{Reduced Dirac Equation for Hydrogenlike Atoms}
\label{sec:equation}

\vskip 0.1cm

In nonrelativistic $QM$, a hydrogenlike atom (shown in Fig.1) is treated by Schr\"{o}dinger equation in the
center-of-mass coordinate system ($CMCS$) as ($\hbar=c=1$)
\begin{equation}\label{shro}
\begin{array}{c}
i\dfrac{\partial}{\partial t}\psi({\bf r}_1,{\bf r}_2,t)=\left[\dfrac{1}{2m}\hat{\bf p}_1^2+\dfrac{1}{2m_N}\hat{\bf p}_2^2-\dfrac{Z\alpha}{r}\right]\psi({\bf r}_1,{\bf r}_2,t) \\[5mm]
=\left[\dfrac{1}{2\mu}\hat{\bf p}^2-\dfrac{Z\alpha}{r}\right]\psi({\bf r},t)
\end{array}
\end{equation}
Here $m=m_e$ and $m_N$ are the masses of electron and nucleus while
\begin{equation}\label{mu}
\mu=\dfrac{mm_N}{m+m_N}\equiv\dfrac{mm_N}{M}
\end{equation}
is the reduced mass. Eq.(\ref{shro}) is formally written down in a relative motion coordinate system ($RMCS$) with the
"point nucleus" being its center and ${\bf r}={\bf r}_1-{\bf r}_2$ ($\hat{\bf p}=-i\hbar\nabla$). Thus a two-body problem is reduced into a one-body problem in a noninertial frame like $RMCS$.

However, for relativistic $QM$ ($RQM$), we cannot bring two kinetic energy terms in the $CMCS$
(an inertial frame) into one like that in Eq.(\ref{shro}) (See the page note after Eq.(\ref{massinv})). When Dirac equation is used for a hydrogenlike atom, one just put $V(r)=-\frac{Z\alpha}{r}$ directly into Eq.(\ref{dirac2}), yielding
\begin{equation}\label{dirac3}
\left\{\begin{array}{l}
(i\hbar\dfrac{\partial}{\partial t}-V(r)-mc^2)\phi=-i\hbar c{\boldsymbol\sigma}\cdot\nabla\chi, \\[4mm]
(i\hbar\dfrac{\partial}{\partial t}-V(r)+mc^2)\chi=-i\hbar c{\boldsymbol\sigma}\cdot\nabla\phi
      \end{array}\right.
\end{equation}
Notice that here two approximations have been made implicitly:\\
(a) The nucleus mass $m_N\to\infty$, so $\mu=m=m_e$ in Eq.(\ref{mu}).\\
(b) The invariance of space-time inversion, Eq.(\ref{inver}) (${\bf x}\to {\bf r}$), must be supplemented by
\begin{equation}\label{vtrans}
V({\bf r},t)\to V(-{\bf r},-t)=-V({\bf r},t)
\end{equation}
even $V(r)$ doesn't contain time $t$ explicitly. \footnotemark[3]
\footnotetext[3]{Previously, the $V$ in Eq.(\ref{vtrans}) was called as a "vector potential", having the same property
like that of "energy" in the Lorentz transformation. Here the physical meaning of $V$ is the electron's potential
energy in an "external field" of nucleus. }

In our point of view, what Eq.(\ref{vtrans}) means is: Under the space-time inversion, while the electron transforms into
a positron, the nucleus remains unchanged at all! Hence the nucleus is treated as an "inert core" in Eq.(\ref{dirac3})
not only in the sense of (a), but also in that of (b).

In a prominent paper \cite{22}, by using Dirac's method, Marsch rigorously solved the hydrogen atom as a two Dirac
particle system bound by Coulomb force. His solutions are composed of positive and negative pairs, corresponding
respectively to hydrogen and antihydrogen as expected. However, surprisingly, in the hydrogen spectrum, besides the
normal type-1 solution with reduced mass $\mu$, there is another anomalous type-2 solution with energy levels:
\begin{equation}\label{}
E'_n=Mc^2-2\mu c^2+\dfrac{1}{2}\mu c^2\left(\frac{\alpha}{n}\right)^2+\cdots,\quad (n=1,2,\ldots)
\end{equation}
And, "strange enough, the type-2 ground state ($n=1$) does not have lowest energy but the continuum ($n=\infty$)"\cite{22}.

In our opinion, these anomalous solutions just imply a positron moving in the field of proton. So all discrete states
with energy $E'_n$ are actually unbound, they should be and can be ruled out in physics by either the "square integrable
condition" or the "orthogonality condition" acting on their rigorous $WF$s (for one body Dirac equation, see \cite{21}, also p.28-31, 50 of \cite{24}).
On the other hand, all continuum states ($n=\infty$) with energies lower than $Mc^2-2\mu c^2$ correspond to scattering $WF$s with negative phase shifts, showing the repulsive force between positron and proton (\cite{25}, section 1.5 in \cite{24} or
section 9.5 in \cite{4}). Marsch's work precisely validates our understanding: (a) The negative energy state
of a particle just describes its antiparticle state. (b) The Coulomb potential allows a complete set of solutions
comprising two symmetric sectors, hydrogen and antihydrogen. In the hydrogen sector, the negative energy states mean that the proton remains unchanged but the electron has already been transformed into a positron under the Coulomb interaction.

However, the nucleus of a hydrogenlike atom maybe either a fermion or a boson like deuteron $d$ (of a deuterium atom $D$) with angular momentum $I=1$. To solve two-particle problem individually would be a daunting task, it couldn't be
rigorous eventually too. We prefer to improve the one-body Dirac equation, Eq.(\ref{dirac3}), at the least labor cost.
It is possible, just let the reduced mass $\mu$ replacing the $m$ in Eq.(\ref{dirac3}) and claim the invarince of
"mass inversion" in a noninertial frame ($RMCS$), ignoring a small centripetal acceleration of the nucleus in the $CMCS$:
\begin{equation}\label{massinv}
\mu\to -\mu,\quad \phi({\bf r},t)\to\chi({\bf r},t),\quad \chi({\bf r},t)\to\phi({\bf r},t)
\end{equation}
Such a reduced Dirac equation ($RDE$) \cite{26} should be tested by experiments. \footnotemark[4]
\footnotetext[4]{Since the nucleus is assumed to be "inert" in the sense of Eq.(\ref{vtrans}), when $m\to -m,m_N\to m_N$,
$\mu \to -\mu(1+\frac{2m}{M})$ (whereas $V(r)$ remains unchanged under the mass inversion). So Eq.(\ref{massinv}) has an
inaccuracy up to $\frac{2m}{M}$ ($<1.1\times 10^{-3}$ for $H$) in this claim.}
Thanks to remarkable advances in high
resolution laser spectroscopy and optical frequency metrology, the $1S-2S$ two-photon transition in atomic hydrogen $H$
(or deuterium $D$) with its natural line width of only 1.3Hz has been measured to a very high precision. In 1997, Udem
\etal,determined the $1S-2S$ energy interval of $H$ being (\cite{27}, see also \cite{43}):
\begin{equation}\label{fh}
f_H^{exp}(1S-2S)=2466061413187.34(84)\,kHz
\end{equation}

In 1998, Huber \etal measured the isotopy-shift of the $1S-2S$ transition of $H$ and $D$ (\cite{28}, see also \cite{29})
\begin{equation}\label{fd}
f_D^{exp}(2S-1S)-f_H^{exp}(2S-1S)=670994334.64(15)\,kHz
\end{equation}
 As expected, the theoretical values calculated from $RDE$ turn out to be \cite{26}
 \begin{equation}\label{}
 \Delta E_{H}^{RDE}(2S-1S)=2.466067984\times 10^{15}\,Hz
\end{equation}
which is only a bit larger than the measured data, Eq.(\ref{fh}), by $3\times 10^{-6}$, and
\begin{equation}\label{}
 \Delta E_{D-H}^{RDE}(2S-1S)=6.7101527879\times 10^{11}\,Hz
\end{equation}
which is larger than that in Eq.(\ref{fd}) by $3\times 10^{-5}$ only.

Further theoretical modifications will bring the discrepancies down to one order of magnitude respectively \cite{26}.
See Appendix $A$.

\section{Negative Mass Paradox and Antigravity}
\label{sec:antigravity}

\vskip 0.1cm

The well-known Newton's gravitation law reads
\begin{equation}\label{ngr}
F(r)=-G\dfrac{m_1m_2}{r^2}
\end{equation}
where $m_1$ and $m_2$ are the gravitational masses of two particles or macroscopic bodies with spherical symmetry. Then
an acute problem arises: can a body have a negative mass? If so, a bizarre phenomenon would occur as discussed by Bondi
\cite{30}, Schiff\cite{31} and Will\cite{32} respectively. Suppose such a body (with mass  $m_1<0$) is brought close
to a normal body (with mass $m_2>0$) and assume the validity of Newton's dynamical law, Eq.(\ref{newt}), together with
\begin{equation}\label{equ}
m_{inert}=m_{grav}
\end{equation}
Then according to Eqs.(\ref{newt}) and (\ref{ngr}), the positive-mass body ($m_2$) would attract the negative-mass body
($m_1$) whereas $m_1$ would repel $m_2$. The pair (a "gravitational dipole") would accelerate itself without outside propulsion. incredible!

The above problem was named as a "negative mass paradox" in \cite{21}. Although a "positive energy theorem" was proved
since 1979, saying that "the total asymptotically determined mass of any isolated body in general relativity ($GR$) must be
non-negative", we believed\cite{21} that a thorough solution to this paradox will tell us much more. As we learn from $RQM$,
the emergence of negative energy is inevitable and is intimately related to the existence of antiparticle. The previous discussion
enables us to establish a working rule: Any theory, either quantum or classical, being
capable of reflecting the equal existence of particle versus antiparticle, must be invariant under a mass inversion ($m\to -m$).

So it is quite natural to generalize Eq.(\ref{ngr}) into \cite{21}:
\begin{equation}\label{ngrm}
F(r)=\pm G\dfrac{m_1m_2}{r^2}
\end{equation}
where the minus sign holds for $m_1$ and $m_2$ (both positive) being both matters or antimatters whereas the plus sign
holds for one of them being antimatter, meaning that matter and antimatter repel each other.

Note: The root cause of "negative mass paradox" is stemming from an incorrect notion that the distinction between $m$ and $-m$ is absolute. But actually, it is merely relative, not absolute.

Now consider a positronium and an ordinary atom. There will be no gravitational force between them, showing $m_{grav}=0$ for
the positronium relative to any matter. But its $m_{inert}\neq0$ due to Einstein's equation: \footnotemark[1]
\footnotetext[1]{Throughout this paper, mass $m$ refers to the "rest mass" as discussed by Okun \cite{42}.}
\begin{equation}\label{}
E_0=m_{inert}c^2=mc^2
\end{equation}
with the rest energy $E_0$ being positive definite. Hence the $GR$'s "equivalence principle" in the (weak) sense of Eq.(\ref{equ}) ceases to be valid in the case of coexistence
of matter and antimatter. Despite of its great success, $GR$ needs some modification to meet the requirement of $SR$. Indeed, let us look at the Einstein field equation ($EFE$) (see \eg, \cite{33}):
\begin{equation}\label{eins}
R_{\mu\nu}-\dfrac{1}{2}g_{\mu\nu}R=8\pi GT_{\mu\nu}
\end{equation}
On the $RHS$ of $EFE$, the energy-momentum tensor ($EMT$) $T_{\mu\nu}$ is proportional to the mass $m$ of matter. Hence
the mass inversion $m\to -m$ will change the $RHS$ of $EFE$ but not its left-hand-side (containing no mass).

To keep $EFE$ invariant under the mass inversion, a generalization is proposed in \cite{21} that
\begin{equation}\label{tch}
T_{\mu\nu}\to T_{\mu\nu}^{eff}=T_{\mu\nu}-T_{\mu\nu}^c
\end{equation}
(as before, the superscript $c$ refers to antimatter). Notice that the form of $EMT$ is the same for both matter and
antimatter. So under the mass inversion, $T_{\mu\nu}\to -T_{\mu\nu}^c$ and $T_{\mu\nu}^c\to -T_{\mu\nu}$. Thus Eq.(\ref{ngrm}) can be derived from Eq.(\ref{eins}) with modification, Eq.(\ref{tch}), in a weak-field approximation (as
described in \cite{33}).

\section{Summary and Discussion}
\label{sec:summary}

\vskip 0.1cm

There are two invariants in the kinematics of $SR$:
\begin{equation}\label{spec1}
c^2(t_1-t_2)^2-({\bf x}_1-{\bf x}_2)^2=c^2(t'_1-t'_2)^2-({\bf x}'_1-{\bf x}'_2)^2 =\text{const} \\
\end{equation}
\begin{equation}\label{spec2}
E^2-c^2{\bf p}^2=m^2c^4
\end{equation}

In hindsight, it seems quite clear that Eq.(\ref{spec1}) is invariant
under the space-time inversion (${\bf x}\to -{\bf x}, t\to -t$) and Eq.(\ref{spec2})
remains invariant under the mass inversion $m\to -m$. However, these two discrete symmetries are deeply rooted at the dynamics of $SR$. Their implication is focused
on one common essence of nature: Everything is in contradiction, \ie, it contains two sides in confrontation inside.
For instance, an electron's $WF$ Eq.(\ref{2spin}), has two components, $\phi$ and $\chi$:
\begin{equation}\label{elec2}
\psi_{e^-}({\bf x},t)\sim \phi\sim\chi\sim \exp\left[\frac{i}{\hbar}({\bf p}\cdot{\bf x}-Et)\right],\qquad
(|\phi|>|\chi|)
\end{equation}
where $\phi$ dominates $\chi$. Under a space-time inversion, Eq.(\ref{inver}), the electron transforms into a positron
with $WF$ being
\begin{equation}\label{posi2}
\psi_{e^+}({\bf x},t)\sim \chi_c\sim\phi_c\sim \exp\left[-\frac{i}{\hbar}({\bf p}\cdot{\bf x}-Et)\right],\qquad
(|\chi_c|>|\phi_c|)
\end{equation}
where $\chi_c$ dominates $\phi_c$. Note that the observed momentum ${\bf p}$ and energy $E\,(>0)$ are the same for
Eqs.(\ref{elec2}) and (\ref{posi2}), using Eqs.(\ref{pe}) and (\ref{pec}) respectively.$^{\footnotemark[5]}$
\footnotetext[5]{Historically, in 1953, Konopinski and Mahmaud \cite{34} wrote down operator relation like
Eq.(\ref{pec}) in a page note while Schwinger's argument in
1958\cite{35} also contains some insight relevant to that in this paper.}
The variations of complex $WF$s of electron and positron at a fixed point, say ${\bf x}=0$, as functions of time $t$ are
depicted on Fig.2.

Comparing Eq.(\ref{psic}) to Eq.(\ref{elec2}) with Eq.(\ref{posi2}), we see that the original "charge conjugation
transformation $C$" could be regarded as "correct" in form but "incorrect" in explanation. So the discovery of $C$
violation in weak interactions could be viewed as a warning that a correct understanding in physics needs rigorous language
in mathematics, careful comparison with experiments and a sound logic in linking them together. If a theory (\eg, the
"hole theory") can only be explained in ordinary language, it would be likely incorrect, or at least missing something
important in the basic concept.

The symmetry between Eqs.(\ref{elec2}) and (\ref{posi2}) (or Eqs.(\ref{pe}) and (\ref{pec})) can also be ascribed to that
of $i$ versus $-i$, showing again the beauty and power of mathematics.

By providing Eq.(\ref{pec}) as a supplement to Eq.(\ref{pe}), the space-time inversion exhibits itself as the essence of $SR$ to show the symmetry of particle versus antiparticle. $SR$ is compatible or in conformity with $QM$ essentially. In some
senses, the reason why $RQM$, $QFT$ and particle physics are capable of developing with vitality is because they have
inherited their "genes" ($DNA$) half (Eq.(\ref{pe})) from $QM$ and half (Eq.(\ref{pec})) from $SR$.

Being different in form but equivalent in essence, the mass inversion seems more convenient in use for implementing the
particle-antiparticle symmetry, especially for a classical theory. The prediction about the antigravity between matter
and antimatter, though interesting, remains open to scientific verification. As now the antihydrogen atoms have already
been made in laboratories on Earth, it's time to consider the universe being filled not only with matter galaxies, but
also with antimatter galaxies at remote distances. A tentative model calculation is currently being studied \cite{36}.

Recently, two experiments have been proposed at Fermilab \cite{37} and CERN \cite{38} respectively, aiming at directly
measuring the free fall acceleration of antihydrogen in the field of Earth (quoted from \cite{39}). We anticipate a big
surprise from such an experiment. We don't believe in the claim that existing experiments already place stringent bounds
(say, $10^{-7}$ \cite{39}) on any gravitational asymmetry between matter and antimatter. This is because an electron in
motion as shown by Eq.(\ref{elec2}) does contain some "hidden positron field" described by $\chi$, which is by no means
a real "positron particle" ingredient with opposite charge. Being subordinate to $\phi$ in an electron, $\chi$ can only
display itself by various $SR$ effects, including $E_0=mc^2$, the time dilatation and the enhancement of
electron's charge at high-energy collisions, \etc ~Similarly, in our opinion, the virtual $e^+e^-$ pairs in the loops of vacuum polarization and self-energy
of $QFT$ (see Figs.1 and 2 in \cite{39}) contain no real antimatter content (of $e^+$) too. While there are many arguments
against "antigravity" \cite{41}, only further experiments can judge.
\section*{Acknowledgements}

\vskip 0.1cm

We thank E. Bodegom, Y. X. Chen, T. P. Cheng, X. X. Dai, Y. Q. Gu, F. Han, J. Jiao, A. Khalil, R. Konenkamp,
D. X. Kong, J. S. Leung, P.T. Leung, D. Lu, Z. Q. Ma, E. J. Sanchez, Z. Y. Shen, Z. Q. Shi, P. Smejtek, X.T. Song,
R. K. Su, F. Wang, Z. M. Xu, J. Yan, R. H. Yu, Y. D. Zhang and W. M. Zhou for encouragement, collaboration and helpful discussions.


\section*{Appendix A: Comparison between Dirac Equation and Reduced Dirac Equation for hydrogenlike atoms}
\begin{tabularx}{165mm}{|p{30mm}|X|p{65mm}|X|p{65mm}|X|}
  \hline
   & Dirac Equation & Reduced Dirac Equation \\
  \hline
  Approximation made & keep $m_e$, \ie, assume $m_N\to\infty$ (combine $CMCS$ and $RMCS$ into one frame)  & keep $m_N$ finite but reduce $m_e$ into $\mu=\frac{m_em_N}{m_e+m_N}=\frac{m_em_N}{M}$ in the $RMCS$ (noninertial frame)  \\
  \hline
   Invariance under space-time inversion & $\begin{array}{c}
                                             {\bf r}_1\to -{\bf r}_1,t\to -t, \\
                                             V({\bf r}_1,t)\to V(-{\bf r}_1,-t)= -V({\bf r}_1,t), \\
                                             \phi({\bf r}_1,t)\to \phi(-{\bf r}_1,-t)= \chi({\bf r}_1,t), \\
                                            \chi({\bf r}_1,t)\to \chi(-{\bf r}_1,-t)= \phi({\bf r}_1,t)
                                           \end{array}$
       & $\begin{array}{c} {\bf r}={\bf r}_1-{\bf r}_2, {\bf r}\to -{\bf r},t\to -t,\\ V({\bf r},t)\to V(-{\bf r},-t)= -V({\bf r},t),\\ \phi({\bf r},t)\to \phi(-{\bf r},-t)= \chi({\bf r},t),\\ \chi({\bf r},t)\to \chi(-{\bf r},-t)= \phi({\bf r},t) \end{array}$ \\
  \hline
   Invariance under mass inversion & $\begin{array}{c} m_e\to -m_e,\\ V({\bf r}_1)\to V({\bf r}_1),\\ \phi({\bf r}_1,t)\to \chi({\bf r}_1,t),\\ \chi({\bf r}_1,t)\to \phi({\bf r}_1,t)\end{array}$ & $\begin{array}{c} m_e\to -m_e, m_N\to m_N, \\ \mu\to -\mu(1+\frac{2m_e}{M})\sim -\mu,\\ V({\bf r})\to V({\bf r}),\\ \phi({\bf r},t)\to \chi({\bf r},t),\\ \chi({\bf r},t)\to \phi({\bf r},t)\end{array}$ \\
  \hline
   Physical implication & With the increase of nucleus charge number $Z$, the electron's energy decreases with $|\chi|$ rising against $|\phi|$. But even when electron turns into a positron ($|\chi|>|\phi|$), the nucleus remains unchanged at all. & Same as the case of Dirac equation even the mass of nucleus is finite.  \\
  \hline
  \vspace*{-2cm}Theoretical prediction \newline (its discrepancy from the experimental data) & $\begin{array}{c}\Delta E^{Dirac}_H(2S-1S)\\ =2.467411048\times 10^{15} Hz\\ (5.5\times 10^{-4}),\\ \Delta E^{Dirac}_{D-H}(2S-1S)=0\\ (100\%)\end{array}$  & $\begin{array}{c}\Delta E^{RDE}_H(2S-1S)\\ =2.466067984\times 10^{15} Hz\\ (3\times 10^{-6}),\\ \Delta E^{RDE}_{D-H}(2S-1S)\\ =6.7101527879\times 10^{11} Hz\\ (3\times 10^{-5}) \end{array}$ \\
  \hline
\end{tabularx}

\section*{Appendix B: Hints from Philosophy}

\vskip 0.1cm

Fig.2 could be compared with Fig.3.


\begin{figure*}[!h]
\centerline{\includegraphics[scale=0.8]{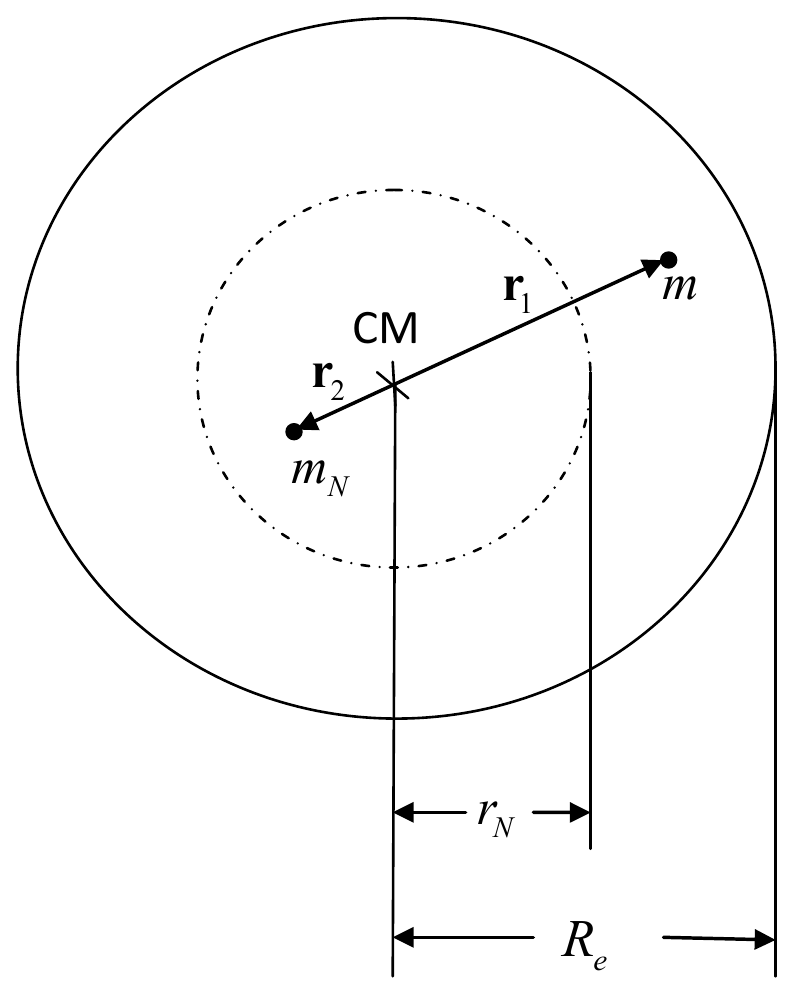}}
\vskip 0.5cm
\caption{
A hydrogenlike atom in quantum mechanical description. The nucleus with mass $m_N$ occupies a small sphere with radius $r_N$
(greatly exaggerated in the diagram) while the electron with mass $m$ spreads over a larger sphere with radius $R_e$
(\ie atomic radius). Their common center is the atom's center of mass (CM). The wavefunction $\psi({\bf r})e^{-iEt}$ with
${\bf r}={\bf r}_1-{\bf r}_2$ shows the electron's amplitude under a "fictitious measurement" \cite{4}, during which the
electron and nucleus shrink into two "fictitious point particles " located at ${\bf r}_1$ and ${\bf r}_2$ simultaneously.
The Coulomb potential $V(r)=-\frac{Ze^2}{r}$ between them is a static one. The probability
to find the electron at ${\bf r}$ is $|\psi({\bf r})|^2$ while that to find its momentum being ${\bf p}$ is $|\phi({\bf p})|^2$ with $\phi({\bf p})$ being the Fourier transform of $\psi({\bf r})$.}
\end{figure*}

\begin{figure*}[!h]
\centerline{\includegraphics[scale=1.0]{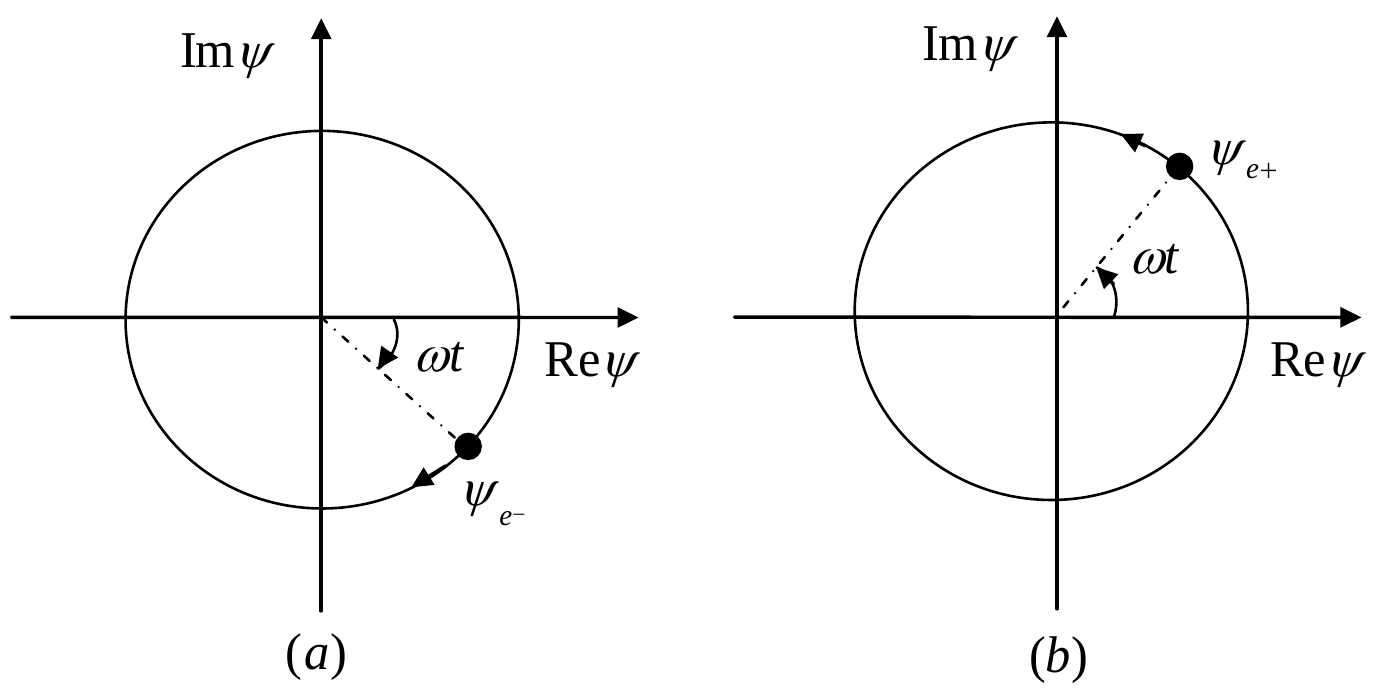}}
\vskip 0.5cm
\caption{The $WFs$ of (a) electron: $\psi_{e^-}(0,t)=e^{-i\omega t}$ and (b) positron: $\psi_{e^+}(0,t)=e^{i\omega t},\;
(\omega=\frac{E}{\hbar})$. Their momentum $p$ is perpendicular to the paper and both particles move towards us ($p>0$). The points $\psi_{e^-}$ and $\psi_{e^+}$ rotate on the unit circle clockwise and anticlockwise respectively. $Re\psi$ and $Im\psi$
transform into each other during the particle's motion or a gauge (phase) transformation. So the distinction between
them is merely relative, not absolute. (see Appendix B) \cite{19}.}
\end{figure*}

\begin{figure*}[!h]
\centerline{\includegraphics[scale=0.7]{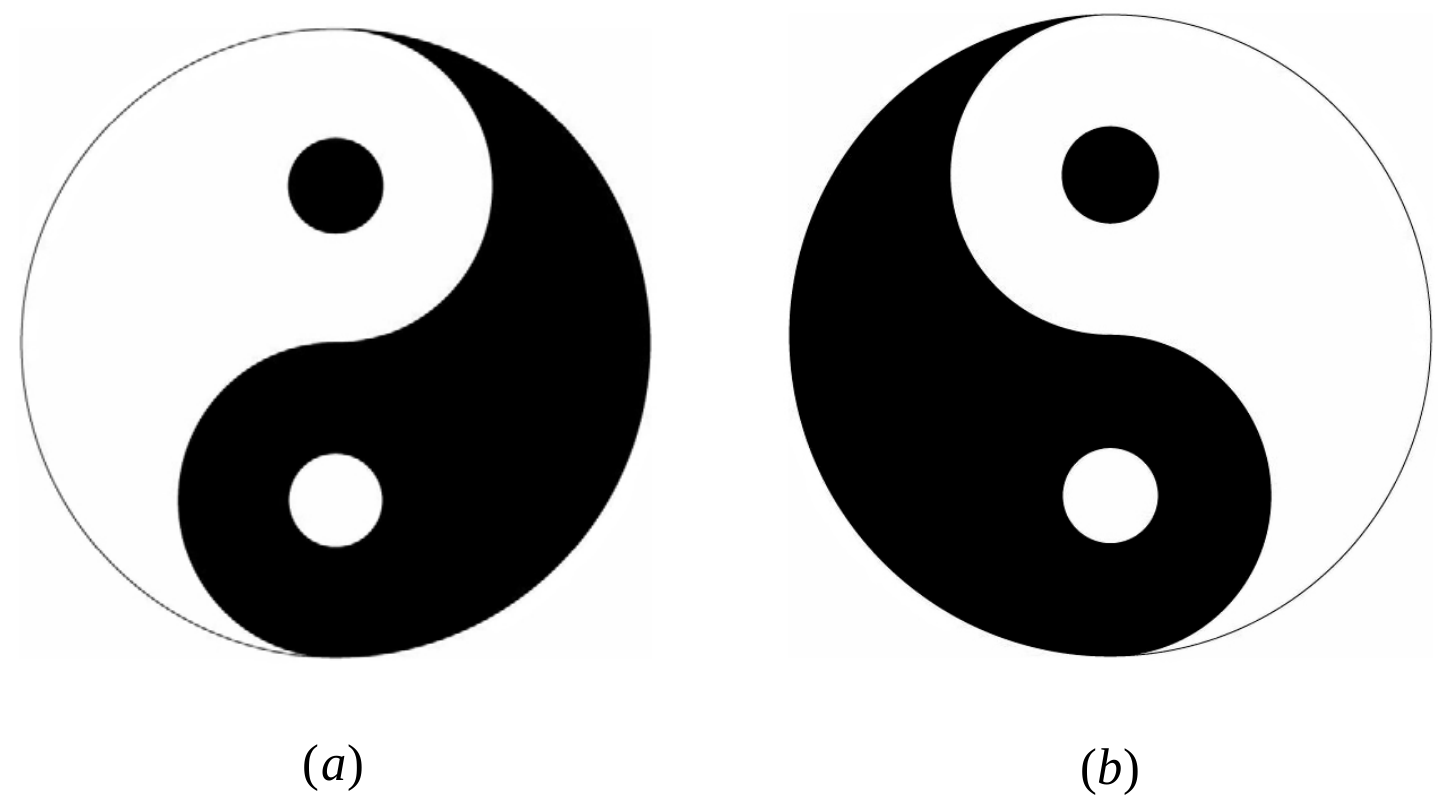}}
\vskip 0.5cm
\caption{Tai-chi Tu (Diagram of the supreme ultimate \cite{40}). (a) The original one and (b) its "mirror image" are rotating clockwise and anticlockwise respectively. Black and white colors refer to "yin" and "yang". Two small circles hiding inside
imply that "there is yang hiding inside yin and vice versa". And the motion is triggered by mutual interactions between yin
and yang inside, not due to a push from the outside. The yin and yang could be corresponding to, of course not precisely,
the $Re\psi$ and $Im\psi$ of a $WF$ $\psi$ (see Fig.2)\cite{19}.
}
\end{figure*}

\end{document}